\documentstyle[12pt]{article}

\def \be {\begin{equation}}
\def \ee {\end{equation}}
\def \ben {\begin{eqnarray}}
\def \een {\end{eqnarray}}
\begin{document}
\title{Scalar and Spinor Particles in the Spacetime of a Domain Wall
in String Theory}
\author{V. B. Bezerra$^{1}$,
L. P. Colatto$^{2}$, M. E. X. Guimar\~aes$^3$ \\
and R. M. Teixeira Filho$^4$ \\
\mbox{\small{1. Departamento de F\'{\i}sica, Universidade Federal da
Para\'{\i}ba}} \\
\mbox{\small{2. Instituto de F\'{\i}sica, Universidade de Bras\'{\i}lia}} \\
\mbox{\small{3. Departamento de Matem\'atica, Universidade de
Bras\'{\i}lia }} \\
\mbox{\small{4. Instituto de F\'{\i}sica, Universidade Federal
da Bahia}} \\
\mbox{\small{\bf valdir@fisica.ufpb.br, colatto@fis.unb.br,}} \\
\mbox{\small{\bf emilia@mat.unb.br , rmuniz@ufba.br}}  \\
\mbox{\small{PACS Numbers: 04.62, 11.25, 98.80C}} \\
\mbox{\small{Keywords: Topological Defects, Scalar-Tensor Gravities,}} \\
\mbox{\small{ Klein-Gordon Equation, Dirac Equation}}}
\maketitle
\date{}
\begin{abstract}
We consider scalar and spinor particles in the spacetime of a domain wall
in the context of low energy effective string theories, such as the
generalized scalar-tensor gravity theories. This class of theories allows
for an arbitrary coupling of the wall and the (gravitational) scalar field.
First, we derive the metric of a wall in the weak-field approximation and
we show
that it depends on the wall's surface energy density and on two
post-Newtonian parameters. Then, we solve the Klein-Gordon
and the Dirac equations in this spacetime. We obtain the spectrum of energy
eigenvalues and the current density in the scalar and spinor cases,
respectively. We show that these quantities, except in the case
of the energy spectrum for a massless spinor particle,  depend on the 
parameters that characterize the scalar-tensor domain wall.
\end{abstract}
\section{Introduction}

Topological defects arise whenever a symmetry is spontaneously broken. They
can be of various types according to the topology of the vacuum manifold
of the field theory being under consideration. In this work, we will
concentrate our attention to domain walls which are defects arising from a
breaking of a discrete symmetry by means of a Higgs
field \cite{kib,vil1,cve1}.

Domain walls have been extensively studied in the litterature. In particular,
it was soon realized that they may lead to a cosmological
catastrophe \cite{vil1}, even if they were produced in a late time phase
transition \cite{hill}. From the gravitational point of view, an
interesting feature of the wall's
gravitational field is that its weak field approximation does not correspond
to any exact static solution of the Einstein's equations, hence implying
that they are gravitationally unstable \cite{vil2}. In the
references \cite{vil3}, a time-dependent metric was obtained and it was shown
that observers experience a repulsion from the wall. Current-carrying walls
and their cosmological consequences were also object of investigations. In
the reference \cite{pat}, the internal structure of a surface
current-carrying wall was studied and the internal
quantities such as energy per
unit surface and the surface current were calculated numerically.

The above mentioned features of a domain wall were analysed in the
framework of the Einstein's theory of gravity. However, it has been
argued that gravity may  be
described by  a scalar-tensorial gravitational field, at least at
sufficiently high energy scales. Indeed, a scalar
field $\phi$, which from now on we will
call generically as {\it dilaton}, appears as a necessary partner of the
graviton field $g_{\mu\nu}$ in all superstring models \cite{sche,dam1}.
Topological defects of various types and their gravitational effects
have already been studied in the framework of various
low energy effective string models \cite{all,mexg}. In what
concernes the domain walls solutions, these configurations were the
object of references \cite{domain}, in which the authors have studied
the properties of the wall's gravitational field  in Brans-Dicke
and in dilatonic gravities. In this class of solutions, the dilaton can 
couple to the matter potential forming the wall. It is shown that the 
dilaton's solution varies with the spatial distance from the wall giving 
rise to a defect called ``dilatonic domain wall".

The aim of this paper is two fold. First, we investigate the gravitational 
field of a domain wall in the context of a
generalized scalar-tensor gravity. Second, we analyse how particles are 
affected by this particular
gravitational field. The gravitational
interaction on quantum mechanical systems has been
studied by many authors \cite{cj}. For this purpose the Klein-Gordon and
Dirac equations in covariant form have been used and solved in curved
spacetimes. The search for these solutions is very interesting under and may
be accounted for by the scheme of unifying quantum mechanics and general
relativity. As examples of works concerning this subject we can mention
Audretsch and Sch\"{a}fer \cite{schafer} who presented a detailed analysis of
the energy spectrum of the hydrogen atom in Robertson-Walker universes and
Parker \cite{parker,parker2} who studied a one-electron atom in a curved
spacetime.

In the present work, we are particularly interested in studying scalar
and spinor
particles in the spacetime of a scalar-tensorial domain wall. In section 2,
we derived the metric of a wall in the weak field approximation. We show
that it depends on the wall's surface energy density $\sigma$ and on two
post-Newtonian parameters, $G_0$ and
$\alpha^2(\phi_0)$. In the section 3, we solve the Klein-Gordon
equation, we find the energy eigenvalues and we point out the dependence
of the current on the parameters that characterize the scalar-tensor 
domain wall. In section 4, we first
consider the Dirac equation for a massive spinor field. Then, we solve
explicitly the Weyl
equations for a massless spinor field and we determine the expression for
the energy spectrum and for the current. Finally, in section 5, we 
present our concluding remarks.

\section{The metric of a domain wall in the weak-field approximation}

In this section, we will derive the metric of a domain wall in a low energy
effective string model, in which the axion field is vanishing. This action
is analogous to the class of scalar-tensor theories developed in the
refs.\cite{wag} in the case of the scalar sector of the gravitational
interaction is massless.
For technical purpose, it is better to work in the so-called
Einstein(conformal) frame in which the kinematic terms of tensor and scalar
fields do not mix.
Then, a domain wall solution arises from the action:
\begin{equation}
\label{3}
S =  \frac{1}{16 \pi G_*} \int
\mbox{d}^4 x \sqrt{-g} \left[ R - 2 g^{\mu \nu} \partial_{\mu} \phi
\partial_{\nu} \phi \right] +
  \int d^4 x \sqrt{-g} A^2(\phi)
\left[ \frac{1}{2}g^{\mu\nu}\partial_{\mu}\Phi \partial_{\nu}\Phi  - V( \Phi)
\right]  ,
\end{equation}
where $g_{\mu \nu}$ is a pure rank-2 metric tensor,
$R$ is the curvature scalar associated to it and $G_*$ is some ``bare"
gravitational coupling constant. The second term in the r.h.s. of eq. (1)
is the matter action representing a model of  a real Higgs scalar
field $\Phi$  and the symmetry breaking potential
 $V( \Phi )$ which possesses a discrete set of degenerate minima.
Action (1) is obtained from the original action
appearing in the refs.  \cite{wag} by a conformal transformation
(see, for instance, \cite{dam})
\begin{equation}
\label{4}
\tilde{g}_{\mu \nu} = A^{2}(\phi) g_{\mu \nu} \;\; ,
\end{equation}
where $\tilde{g}_{\mu\nu}$
is the physical metric which contains both  scalar and tensor degrees of
freedom,
and by a redefinition of the quantities
\[
G_*A^{2}(\phi) = \frac{1}{\tilde{\Phi}} \;\; ,
\]
where $\tilde{\Phi}$ is the original scalar field, and
\[
\alpha(\phi) \equiv \frac{\partial \ln A}{\partial \phi} =
\frac{1}{[2 \omega(\tilde{\Phi}) + 3]^\frac{1}{2}}\;\; ,
\]
which can be interpreted as the (field-dependent) coupling strenght
between matter and the scalar field. We choose to leave $A^2(\phi)$
as an arbitrary function of the dilaton field.

In the Einstein frame, the field equations are written as follows:
\begin{eqnarray}
\label{ee}
R_{\mu\nu} & = &  2\partial_{\mu}\phi\partial_{\nu}\phi +
8\pi G_*(T_{\mu\nu} -
\frac{1}{2}g_{\mu\nu}T) \nonumber \\
& & \Box_{g}\phi = - 4\pi G_* \alpha(\phi) T
\end{eqnarray}
where the energy-momentum tensor is obtained as
\[
T_{\mu\nu} \equiv \frac{2}{\sqrt{-g}} \frac{\delta S_m}{\delta g_{\mu\nu}} .
\]

In what follows, we will consider the solution of a domain wall in the
$yz$-plane in the weak-field approximation. Therefore, we will expand
eqs. (\ref{ee}) to first order in $G_* A^2 (\phi_0)$ in such a way that
\begin{eqnarray}
g_{\mu\nu} & = & \eta_{\mu\nu} + h_{\mu\nu} \nonumber \\
\phi & = & \phi_0 + \phi_{(1)} \\
A(\phi) & = & A(\phi_0)[ 1 + \alpha(\phi_0)\phi_{(1)}] \nonumber \\
T^{\mu}_{\nu} & = & T^{\mu}_{(0)\nu} + T^{\mu}_{(1)\nu} \nonumber
\end{eqnarray}
In this approximation, $T^{\mu}_{(0)\nu} =
A^2(\phi_0)\tilde{T}^{\mu}_{(0)\nu}$ is the energy-momentum tensor
of a static domain wall with neglegible width and lying in a $yz$-plane.
Therefore,
\begin{equation}
\label{dw}
T^{\mu}_{(0)\nu} = A^2(\phi_0)\sigma \delta(x) diag (1,0,1,1)
\end{equation}
in the cartesian coordinate system $(t,x,y,z)$.  The parameter $\sigma$ 
is the wall's surface energy density. In our convention, the metric 
signature is $-2$.

The equations (\ref{ee}) in the linearised regime reduce to:
\begin{eqnarray}
\label{li}
\nabla^2 h_{\mu\nu} & = & 16\pi G_* (T_{(0)\mu\nu} -
\frac{1}{2} \eta_{\mu\nu}T_{(0)}) \\
\nabla^2 \phi_{(1)} & = & 4\pi G_* \alpha(\phi_0)T_{(0)} \nonumber
\end{eqnarray}

Let us begin by solving the equation for the dilaton field $\phi_{(1)}$
in (\ref{li}):
\begin{eqnarray}
\nabla^2\phi_{(1)} & = & 12\pi \sigma G_0\alpha(\phi_0)\delta(x) \nonumber \\
\phi_{(1)} & = & 6 \pi \sigma G_0\alpha(\phi_0)\mid x\mid ,
\end{eqnarray}
where $G_0 \equiv G_* A^2(\phi_0)$.

Now, the linearised Einstein's equation in (\ref{li}) with source given by
(\ref{dw}) are just the same as in Vilenkin's paper\cite{vil2}, except that
in our case the metric is multiplied by the factor $A^2(\phi)$ linearised.
Therefore, we have (to first order in $G_0$):
\begin{equation}
\label{VBB}
ds^2 = A^2(\phi_0) \left[ 1 +
4\pi \sigma G_0  |x|(3 \alpha^2 (\phi_0) - 1) \right] [dt^2 - dx^2 - dy^2 -
dz^2] .
\end{equation}
The factor $A^2(\phi_0)$ appearing in the above expression can be absorbed by
a redefinition of the coordinates $(t,x,y,z)$. We finally, then, obtain:
\begin{equation}
\label{me}
ds^2 =  ( 1 + 4D|x|)
[dt^2 - dx^2 - dy^2 - dz^2] .
\end{equation}
where $D\equiv\pi\sigma G_0(3\alpha^2-1)$.

This is the line element corresponding to a domain wall in the framework
of scalar-tensor gravity in the weak-field approximation. The geometry
given by eq.(\ref{me}) is only valid for $Dx\ll 1$.

\section{Klein-Gordon equation in scalar-tensor domain wall}

Let us consider a scalar quantum particle embedded in a
classical background
gravitational field. Its behavior is described by the covariant
Klein-Gordon equation
 \begin{equation}
\label{KG}
\left[ \frac{1} {\sqrt{-g}} \partial_\mu(\sqrt{-g} g^{\mu\nu}\partial_{\nu})
 + m^2\right] \psi = 0,
\end{equation}
where $m$ is the mass of the particle, $g$ is the determinant of the metric
tensor $g_{\mu\nu}$ and we are considering a minimal coupling.

In the space-time of a scalar-tensor domain wall given by
metric (\ref{me}), eq.(\ref{KG}) takes the form
\begin{equation}
\left[ \frac{1}{1+4D|x|} \left(\partial_{t}^{2}-\partial_{x}^{2}-
\partial_{y}^{2}-
\partial_{z}^{2}-\frac{4D} {1+4D|x|} \frac{d|x|}
{dx}\partial_{x}\right) + m^2\right]\psi=0.
\end{equation}
Multiplying this equation
by  $1+4D|x|$ and neglecting terms of order
$D^2$ and up (because we are working in the weak field aproximation), we get
\begin{equation}
\label{KG_W}
\left[\partial_{t}^{2}-\partial_{x}^{2}-\partial_{y}^{2}-\partial_{z}^{2}-
4D\frac{d|x|}
{dx}\partial_{x}+m^2(1+4D|x|)\right]\psi=0.
\end{equation}

Since eq. (\ref{KG_W}) is invariant under the transformation
$x\rightarrow -x$,
we shall restrict the allowed values of $x$ to the interval $x>0$. Then,
we have
\begin{equation}
\label{KG_W*}
\left[\partial_{t}^{2}-\partial_{x}^{2}-\partial_{y}^{2}-\partial_{z}^{2}-
4D\partial_x+m^2(1+4Dx)\right]\psi=0.
\end{equation}

Let us assume that
\begin{equation}
\label{phi}
\psi (t,x,y,z)= e^{-i(Et-k_y y-k_z z)}X(x),
\end{equation}
where $E$, $k_y$ and $k_z$ are constants. If we substitute
relation (\ref{phi}) into eq. (\ref{KG}), we obtain
\be
\label{EqX}
\frac{d^2X(x)}{dx^2}+4D\frac{dX(x)}{dx}+\left[E^2-k_y^2-k_z^2-m^2(1+
4Dx)\right]X(x)=0,
\ee
whose general solution is
\ben
\label{X1}
&&
X(x) = \left\{ C_1AiryAi\left[-(-1)^{1/3}2^{-4/3}(m^2D)^{-2/3}(m^2-E^2+
k_y^2+k_z^2\nonumber \right. \right.\\
&&
\phantom{aaaaaa} \left. \left.  + 4m^2Dx+4D^2)\right]+
C_2AiryBi\left[-(-1)^{1/3}2^{-4/3}(m^2D)^{-2/3}\right. \right. \nonumber \\
&& \phantom{aaaaaa} \left. \left.(m^2-E^2+k_y^2+k_z^2 +
4m^2Dx+4D^2)\right]\right\} e^{-2Dx}, \een with functions
$AiryAi(x)$ and $AiryBi(x)$ being the Airy functions and
$C_1$ and $C_2$ are integration constants. 
Note that the arguments of the 
$Airy$ functions has only terms of order less than $D^2$. Neglecting terms
of order $\ge D^2$, we get 
\ben
\label{X11} && X(x) \simeq \left\{
C_1AiryAi\left[-(-1)^{1/3}2^{-4/3}(m^2D)^{-2/3}(m^2-E^2+
k_y^2+k_z^2\nonumber \right. \right.\\
&& \phantom{aaaaaa} \left. \left.  + 4m^2Dx+4D^2)\right]+
C_2AiryBi\left[-(-1)^{1/3}2^{-4/3}(m^2D)^{-2/3}\right. \right. \nonumber \\
&& \phantom{aaaaaa} \left. \left.(m^2-E^2+k_y^2+k_z^2 +
4m^2Dx+4D^2)\right]\right\} (1-2Dx), \een 

In order to determine
the bound states energies we must require periodicity conditions in
directions $\emph y$ and $\emph z$, with periods $L_y$ e $L_z$,
respectively, supplemented by boundary conditions that the
solution vanishes at $x=a$ and $x=b$, with $a<b$ and
such that $Da\ll 1$ and $Db\ll 1$. These boundary conditions
can be expressible as 

\ben \label{cc} &&
\psi (t,x,y,z)=\psi(t,x,y+L_y,z) \nonumber \\
&& \psi (t,x,y,z)=\psi(t,x,y,z+L_z) \nonumber \\
&& \psi(t,a,y,z)=\psi(t,b,y,z)=0.
\een
These boundary conditions lead us to the following results
\ben
\label{X2}
&&
X(x) = C_1\left\{AiryAi\left[-(-1)^{1/3}2^{-4/3}(m^2D)^{-2/3}(m^2-E^2+
k_y^2+k_z^2 \nonumber \right. \right.  \\
&&
 \left. \left. +4m^2Dx+4D^2)\right] \right. \nonumber  \\
&&
\left. - \frac{AiryAi\left[-(-1)^{1/3}2^{-4/3}(m^2D)^{-2/3}(m^2-E^2+k_y^2+
k_z^2 + 4m^2Da+4D^2)\right]}{AiryBi\left[-(-1)^{1/3}2^{-4/3}(m^2D)^{-2/3}
(m^2-E^2+k_y^2+k_z^2 + 4m^2Da+4D^2)\right]}\right.  \nonumber   \\
&&
\left.
\times AiryBi \left[-(-1)^{1/3}2^{-4/3}(m^2D)^{-2/3} \right. \right.
\phantom{aaaaaaaaaaa} \nonumber   \\
&&
 \left. \left.  \times (m^2 -E^2+k_y^2+k_z^2 + 4m^2Dx+4D^2) \right]
\right\}(1-2Dx), \een where \ben &&
k_y=\frac{2\pi n_y}{L_y} \phantom{aaaaaaaaaaaa} n_y=0,\pm 1,\pm 2, ...\\
&&
k_z=\frac{2\pi n_z}{L_z} \phantom{aaaaaaaaaaaa} n_z=0,\pm 1,\pm 2, ...\;\;,
\een
and
\be
\label{Xb}
X(b)=0 \;\;.
\ee

The boundary condition given by eq. (\ref{Xb}) determines the energy levels of
the particle in the stationary state in the region under consideration. It
should be stressed that in order to solve this problem we must impose that
$|m^2-E^2+k_y^2+k_z^2+4m^2Dx|\gg |(m^2D)^{2/3}|$, in which case the absolute
value of the argument  of Airy's functions are much greater than unity,
which allows us to use the following assymptotic expansions \cite{abra}
\ben
&&
AiryAi(z)\sim \frac{1}{2\sqrt{\pi}}z^{-1/4}e^{-\xi}\sum^{\infty}_{k=0}
(-1)^kc_k\xi^{-k} \nonumber \\
&&
AiryBi(z)\sim \frac{1}{\sqrt{\pi}}z^{-1/4}e^{\xi}\sum^{\infty}_{k=0}
c_k\xi^{-k}
\een
where
\ben
&&
\xi=\frac{2}{3}z^{3/2} \nonumber \\
&&
c_0=1,\phantom{aaaa} c_k=\frac{(2k+1)(2k+3)...(6k-1)}{216^kk!}
\een
In this aproximation, eq. (\ref{Xb}) can be written as
\ben
e^{-\frac{2}{3}W^{3/2}(b)}\sum^{\infty}_{k=0}(-1)^kc_k{\left[\frac{2}{3}
W^{3/2}(b)\right]}^{-k}- e^{-\frac{4}{3}W^{3/2}(a)}
\phantom{aaaaaaaaaaaaaaaa} &&\nonumber \\
\phantom{aaa} \times
\frac{\sum^{\infty}_{k=0}(-1)^kc_k{\left[\frac{2}{3}W^{3/2}(a)\right]}^
{-k}}{\sum^{\infty}_{k=0}c_k{\left[\frac{2}{3}W^{3/2}(a)\right]}^{-k}}
e^{\frac{2}{3}W^{3/2}(b)}\sum^{\infty}_{k=0}c_k{\left[\frac{2}{3}W^{3/2}(b)
\right]}^{-k}=0, && \een where $W(x) = -
(-1)^{1/3}2^{-4/3}(m^2D)^{-2/3}(m^2 - E^2+k_y^2+k_z^2 +
4m^{2}Dx+4D^2)$.

Considering only the first two terms of the summation and
neglecting terms of order $\ge D^2$, we find \be \label{24}
e^{i2\sqrt{E^2-k_y^2-k_z^2-m^2}(b-a)}\left[1-i\frac{4m^2D(b^2-a^2)}
{\sqrt{E^2-k_y^2-k_z^2-m^2}}\right]=1. \ee Now, let us take
$E^2-k_y^2-k_z^2-m^2>0$, then we can rewrite eq.(\ref{24}) as a
system of equations involving its real and imaginary parts as
follows \ben \label{Re}
1+\frac{4m^2D(b^2-a^2)}{\sqrt{E^2-k_y^2-k_z^2-m^2}}\tan
\left[ 2\sqrt{E^2-k_y^2-k_z^2-m^2}\right]   (b-a) \nonumber && \\
 \phantom{aaaaaaaaaaaaaaaaaaa}
=\sec \left[2\sqrt {E^2-k_y^2-k_z^2-m^2}\right] (b-a) && \een \be
\label{Im} \tan \left[2\sqrt{E^2-k_y^2-k_z^2-m^2}\right](b-a)=
\frac{4m^2D(b^2-a^2)}{\sqrt{E^2-k_y^2-k_z^2-m^2}}. \ee

As the function $\tan(x)$ is of order $D$, we can make the assumption that
$\tan(x)\simeq x $. Then, eq. (\ref{Im}) results in
\be
2\sqrt{E^2-k_y^2-k_z^2-m^2}(b-a)\simeq \frac{4m^2D(b^2-a^2)}
{\sqrt{E^2-k_y^2-k_z^2-m^2}} +n\pi, \phantom{a} n=0,\pm 1,\pm 2,...
\ee
Equation(\ref{Re}) is assured if $n$ is even. Then, we have
\be
\label{28}
2\sqrt{E^2-k_y^2-k_z^2-m^2}(b-a)\simeq \frac{4m^2D(b^2-a^2)}
{\sqrt{E^2-k_y^2-k_z^2-m^2}} +2n\pi,\phantom{a} n=0,\pm 1,\pm 2,...\;\; .
\ee

From previous equation we get, finally, that
\be
E^2=m^2+k_y^2+k_z^2+\frac{n^2\pi^2}{(b-a)^2}+(2+\frac{n\pi}{b-a})
m^2D(b+a), \phantom{aaaaaa} n=1,2,...
\ee

It is worth to note that the presence of the wall increases the energy
eigenvalues with parameters\footnote{Just as a reminder for the reader,
$\sigma , \alpha(\phi_0)$ and $G_0$ are the wall's surface energy density,
the coupling strength between the wall and the dilaton, and the effective
gravitational constant, respectively.} $\sigma$, $\alpha(\phi_0)$ and $G_0$
and that for $D=0$
(absence of the wall) we recover the result corresponding to the Minkowski
spacetime as it should be.

In which concerns the current associated with the scalar field
given by
\begin{equation}
\label{VBB1}
J^{\mu}=\frac{i \sqrt{-g}}{2m}( \Psi^{*} \partial ^{\mu} \Psi
- \Psi \partial ^{\mu} \Psi^{*}),
\end{equation}
it is clear that it depends on the parameters that characterizes
the scalar-tensor domain wall through $\sqrt{- g}$ and the solution $\Psi$
of the Klein-Gordon equation.

\section{Dirac equation in a scalar-tensor domain wall}

Now, let us consider a spinor particle embedded in a classical
gravitational field. The covariant Dirac equation governing the
particle, in a curved spacetime, for a spinor $\Psi$, may be
written as \be \label{Dirac}
[i\gamma^{\mu}(x)\partial_{\mu}+i\gamma^{\mu}(x)\Gamma_{\mu}-m]\Psi(x)=0,
\ee where $\gamma^{\mu}(x)$ are the generalized Dirac matrices and
are given in terms of the standard flat space Dirac matrices
$(\gamma^{(a)})$ as \be
\gamma^{\mu}(x)=e^{\mu}_{(a)}(x)\gamma^{(a)}, \ee where
$e^{\mu}_{(a)}(x)$ are tetrad components defined by \be
e^{\mu}_{(a)}(x)e^{\nu}_{(b)}(x)\eta^{(a)(b)}=g^{\mu\nu}. 
\ee 
The product $\gamma^{\mu}(x)\Gamma_{\mu}$ that appears in Dirac
equation can be written as \be \label{AB}
\gamma^{\mu}(x)\Gamma_{\mu}=\gamma^{(a)}(A_{(a)}(x)+i\gamma^{(5)}B_{(a)}(x)),
\ee with
$\gamma^{(5)}=i\gamma^{(0)}\gamma^{(1)}\gamma^{(2)}\gamma^{(3)}$
and $A_{(a)}$ and $B_{(a)}$ are given by \be
A_{(a)}=\frac{1}{2}(\partial_{\mu}e^{\mu}_{(a)}+e^{\rho}_{(a)}\Gamma_
{\rho\mu}^{\mu}) \ee and \be
B_{(a)}=\frac{1}{2}\epsilon_{(a)(b)(c)(d)}e^{(b)\mu}e^{(c)\nu}
\partial_{\mu}e^{(d)}_{\nu},
\ee
where $\epsilon_{(a)(b)(c)(d)}$ is the completely antisymmetric
fourth-order unit tensor.

In the space-time of a scalar-tensor domain wall given by metric
(\ref{me}), let us choose the following set of tetrads \be
\label{gamma0} e^{(a)}_{\mu}=[1+4D|x|]^{1/2}\delta^a_{\mu}, \ee
which implies that \be \label{gamma} \gamma^{\mu}=
[1-2D|x|]\gamma^{(\mu)}, \ee  in which we have neglected terms of
order $\ge D^2$.

Computing the expressions for $A_{(a)}$ and $B_{(a)}$ and putting
these results into eq. (\ref{AB}) and
neglecting terms of order $\ge D^2$,  we get 

\be \label{gG}
\gamma^{\mu}(x)\Gamma_{\mu} = 3 D \gamma^{(1)}. 
\ee 
Now, using
eqs.(\ref{gamma}) and (\ref{gG}), the Dirac equation 
(\ref{Dirac}) in the spacetime of a scalar-tensor domain wall
reads

\be \label{Dirac1} \left\{ i\gamma^{(\mu)}\partial_{\mu}+i3D
\gamma^{(1)}-[1+2Dx]m\right\} \Psi(x)=0, \ee in which we neglected
terms of order $\ge D^2$ and considered only the interval $x>0$.
In order to determine the solutions in the spinoral case, let us
choose the following representation of Dirac matrices \be
\gamma^{(0)}=\left(\begin{array}{cc}1&0\\0&-1\end{array}\right),
\phantom{aaaa}\gamma^{(i)}=\left(\begin{array}{cc}0&\sigma_i\\-\sigma_i&0
\end{array}\right), \phantom{aaaa} i=1,2,3
\ee
where $\sigma_i\;\;(i=1,2,3)$ are the usual Pauli matrices.

Since we are interested in the qualitative behavior of the
particle with respect to the parameters that define the wall, we
will simplify our analysis considering the solution of the Dirac
equation corresponding to massless spinor particle in which case
eq.(\ref{Dirac1}) reduces to \be \label{Weyl} \left\{
i\gamma^{(\mu)}\partial_{\mu}+i3D\gamma^{(1)}\right\} \Psi(x)=0,
\ee and is supplemented by the helicity condition \be \label{hel}
(1+\gamma^5)\Psi(x)=0, \ee where
$\gamma^5=\left(\begin{array}{cc}0&1\\-1&0\end{array}\right)$.
Equations (\ref{Weyl}) and (\ref{hel}) are the Weyl equations for
a massless spin-$\frac{1}{2}$ particle.

The condition given by ($\ref{hel}$) implies that the four-spinor
$\Psi(x)$ is
such that $\Psi(x)=  \left(\begin{array}{c}\Psi_1(x)\\
\Psi_2(x)\end{array}\right)$, with $\Psi_1(x)=\Psi_2(x)$.

A suitable set of solutions of Weyl's equations is of the form \be
\label{psi} \Psi (t,x,y,z)= \left(\begin{array}{c}{\bf u}(x)\\
{\bf u}(x)\end{array}\right) e^{-i(Et-k_yy-k_zz)}, \ee where ${\bf
u}(x)=\left(\begin{array}{c}u_1(x)\\ u_2(x)\end{array}\right)$. If
we substitute relation (\ref{psi}) into eq. (\ref{Weyl}), we
obtain \ben \label{sys} &&
i\frac{du_1(x)}{dx}-(k_y-3D)u_1(x)+(E+k_z)u_2(x)=0 \nonumber \\
&&
i\frac{du_2(x)}{dx}+(k_y+3D)u_2(x)+(E-k_z)u_1(x)=0.
\een

Neglecting terms of order $D^2$ and up, the set of equations (\ref{sys}) can
be written as
\ben
\label{u}
&&
\frac{d^2u_1(x)}{dx^2}+6D\frac{du_1(x)}{dx}+(E-k_y^2-k_z^2)u_1(x)=0,
\nonumber \\
&&
\frac{d^2u_2(x)}{dx^2}+6D\frac{du_2(x)}{dx}+(E-k_y^2-k_z^2)u_2(x)=0,
\een whose general solution reads \be {\bf u}(x)={\bf C_1}
e^{(-3D+i\sqrt{E^2-k_y^2-k_z^2})x}+{\bf C_2}
e^{(-3D-i\sqrt{E^2-k_y^2-k_z^2})x}, \ee where ${\bf C_1}$ e ${\bf
C_2}$ are constant bispinors. Neglecting terms of order $D^2$
and up, we get 

\be \label{u} {\bf u}(x)=(1-3Dx)({\bf C_1}
e^{i\sqrt{E^2-k_y^2-k_z^2}x}+{\bf C_2}
e^{-i\sqrt{E^2-k_y^2-k_z^2}x}). \ee  Again, we notice that the
solutions depend on $\sigma$, $\alpha$ and $G_0$, as expected.

Now, let us compute the current $j^{\mu}$, which is defined by \be
\label{jmu} j^{\mu}=\overline{\Psi}\gamma^{\mu}\Psi. \ee 

Using eq. (\ref{gamma}), the expression for the current in the approximation 
considered  turns into

\be
\label{VBB2}
j^\mu = (1-4Dx)\Psi^\dagger\gamma^{(0)}\gamma^{(\mu)}\Psi, 
\ee
Substituting 
eqs. (\ref{psi}) and (\ref{u}) into eq. (\ref{VBB2}) and considering, for 
simplicity the case in which ${\bf C_2=0}$, we get
 
\be 
j^\mu = (1-10Dx)({\bf C_1^\dagger}\hspace{0.3cm}{\bf C_1^\dagger})
\gamma^{(0)}\gamma^{(\mu)}
\left(\begin{array}{c} {\bf C_1} \\ {\bf C_1} \end{array}\right).
\ee

In order to obtain the energy spectrum, let us consider the same
boundary conditions given by eq.(\ref{cc}) for the scalar field.
Thus, we have 

\ben \label{ccpsi} &&
\Psi (t,x,y,z)=\Psi(t,x,y+L_y,z) \nonumber \\
&& \Psi (t,x,y,z)=\Psi(t,x,y,z+L_z) \nonumber \\
&& \Psi(t,a,y,z)=\Psi(t,b,y,z)=0. \een 
Analogously to the case of a scalar field, from these boundary 
conditions we get

\be
\label{u2} {\bf u}(x) = {\bf
C_1}e^{i\sqrt{E^2-k_y^2-k_z^2}x}(1-e^{-2i\sqrt{E^2-k_y^2-k_z^2}(x-a)})
\ee where \ben &&
k_y=\frac{2\pi n_y}{L_y} \phantom{aaaaaaaaaaaa} n_y=0,\pm 1,\pm 2, ...\\
&& k_z=\frac{2\pi n_z}{L_z} \phantom{aaaaaaaaaaaa} n_z=0,\pm 1,\pm
2, ...\;\;, \een and \be \label{ub} {\bf u}(b)=0 \;\;. \ee 
The energy levels arrives from eq. (\ref{ub}) and are given by 
\be
E^2=k_y^2+k_z^2+\frac{n^2\pi^2}{(b-a)^2}. 
\ee

Note that the energy spectrum is the same as in the flat Minkowski spacetime
case. This result comes from the fact that the spinor field is massless. The
same coincidence occurs in the case of a massless scalar field.

From previous results we conclude that the current differs from that in
the Minkowski spacetime by terms containing the parameters $\sigma$ ,
$\alpha$ and $G_0$ used to describe the scalar-tensor domain wall and 
tends to the corresponding result in Minkowski spacetime in the absence of 
the wall as it should be.

\section{Concluding Remarks}

Recently there has been a growing interest on domain walls as brane world
scenarios and also on scalar-tensor theories of gravity due to its possible
role in the understanding of the physics of the early Universe when
topological defects like domain walls were formed. At that time the dilaton 
fields as well as the topological defect such as a scalar-tensor domain wall
were, certainly, very relevant. These points
constitute the main motivation for this work.

A scalar or spinor particle placed in the spacetime of a scalar-tensor 
domain wall is perturbed by this background due to the geometrical 
and topological features 
of the spacetime under consideration. In other words,
the dynamic of atomic systems is determined by the curvature at the position
of the system and also by the topology of the background spacetime.

Summarizing our conclusions we can say that the metric of a scalar-tensor
domain wall depends on the wall's surface energy density $\sigma$ and on
two Post-Newtonian parameters  $\alpha(\phi_{0})$ and $G_0$. 
The solutions for the scalar and spinor cases differ from the flat Minkowski
spacetime case by the presence of these parameters. The presence of the wall
shift the energy levels and alters the current in the scalar case as
compared with the flat spacetime. In the massless spinor particle case,
there is no shift in the energy spectrum, but the current is altered by
the presence of the scalar-tensor domain wall.

Finaly, it is worth commenting that the study of a quantum system in a
gravitational field like, for example, the one considered in this paper,
may shed some light on the problems of combining quantum mechanics and
gravity. On the other hand, the investigation of topological defects in the
framework of general scalar-tensor theories seems to be important in order
to understand the role played by these strutures in this general context.

\section{Ackowledgments}
V.B.B. would like to thank CNPq for partial financial support. L.P.C.
would like to thank CAPES
for financial support. L.P.C. and M.E.X.G. acknowledge the
kind hospitality of the {\it ``Grupo de F\'{\i}sica Te\'orica"} of the
Universidade Cat\'olica de Petr\'opolis where part of this work has been done.

\end{document}